\documentclass[twocolumn,preprintnumbers,amsmath,amssymb]{revtex4}
\usepackage{graphicx} % include figure files
\usepackage{setspace} % include figure files
\usepackage{dcolumn} % align table columns on decimal point
\usepackage{bm} % bold math
\usepackage{natbib}
\usepackage{hyperref}
\hypersetup{colorlinks=true,linkcolor=blue,citecolor=blue,urlcolor=blue}
\usepackage{epsfig}
\usepackage{amsmath}
\begin{document}
\title{Configurational Entropy–Driven Phase Stability and Thermal Transport in Rock-Salt High-Entropy Oxides}
\author{Ashutosh Kumar\footnote{Email: ashutosh@iitbhilai.ac.in}}
\author{Adrien Moll}
\author{Jitendra Kumar}
\author{Diana Dragoe}
\author{David Bérardan\footnote{Email: david.berardan@universite-paris-saclay.fr}}
\author{Nita Dragoe\footnote{Email: nita.dragoe@universite-paris-saclay.fr}}

\affiliation{ICMMO (UMR CNRS 8182), Université Paris-Saclay, F-91405 Orsay, France}
\affiliation{Functional Materials Laboratory, Department of Materials Science and Metallurgical Engineering, Indian Institute of Technology Bhilai, Chhattisgarh India}
\date{\today}
\begin{abstract}
High-entropy oxides (HEOs) offer a unique platform for exploring the thermodynamic interaction between configurational entropy and enthalpy in stabilizing complex solid solutions. In this study, a series of rock-salt structured oxides with varying configurational entropy, ranging from binary to multi-cation systems, to elucidate the competing roles of enthalpy and entropy in phase stabilization is investigated. Compositions including (Ni$_{0.8}$Cu$_{0.2}$)O to(NiCuZnCoMg)$_{0.9}$A$_{0.1}$O (A = Li, Na, K) were synthesized and their stuctural, microstructural and thermal properties have been discussed. X-ray diffraction combined with thermal cycling confirms that even a medium configurational entropy ($\sim$ 0.95R) can induce single-phase behavior stabilized by configurational entropy ($\Delta S_{conf}$), challenging the traditional threshold of $1.5\,R$. High-resolution TEM and EDS mapping reveal nanocrytalline features and homogeneous elemental distribution respectively, while XPS analysis confirms divalent oxidation states. A strong coupling between high configurational entropy with thermal conductivity ($\kappa$) has been observed. First, a sharp decrease in $\kappa$ with increasing $\Delta S_{conf}$ is seen and then decomposed samples (while cooling) show high $\kappa$, demonstrating the role of $\Delta S_{conf}$ on $\kappa$. Furthermore, Li-doped compositions exhibit improved thermoelectric performance, with a maximum figure of merit ($zT$) of $\sim$0.15 at 1173K\, driven by low thermal conductivity and favorable carrier transport. The results highlight that configurational entropy, even at intermediate values, plays a significant role in stabilizing disordered single-phase oxides and tailoring phonon transport.
\end{abstract}
\maketitle
\section{Introduction}
Thermodynamic phase stability driven by configurational entropy has the potential to enable complex solid solutions composed of several main elements. The concept of high entropy was applied originally in metal\cite{R1} which has been expanded in ionic materials\cite{R2,R3}, semiconductors\cite{R4}, low-dimensional materials\cite{R5} etc. Rost et al. demonstrated the development of a new composition using the high configurational entropy approach labeled as entropy-stabilized oxide (Mg$_{0.2}$Co$_{0.2}$Ni$_{0.2}$Cu$_{0.2}$Zn$_{0.2}$)O with a rock-salt structure, coining the impetus of ionic bonded high entropy oxide materials\cite{R2}. Mixing multiple elements in equimolar ratio results in a high configurational entropy which can give rise to novel forms of random or loosely ordered clusters. The configurational entropy is defined as  $\Delta S_{conf}=-R\sum_{i=1}^n x_ilnx_i$. The systems with $\Delta S_{\mathrm{conf}} \geq 1.61R$ are termed as high-entropy systems, however, when the thermodynamic phase stability is primarily influenced by high configurational entropy $(\Delta S_{conf}$), it is known as an entropy-stabilized system that is further confirmed via a reversible transformation from a multiphase to a single-phase system with the change in sintering temperature above and below the critical temperature.\cite{R2} The vast range of atomic configurations presents opportunities to discover new functionalities, along with unique local symmetries, ordering behaviors, and interstitial arrangements.\cite{R5a} Several interesting physical properties have been demonstrated in seminal compositions derived from the first entropy-stabilized oxide (ESO) following its discovery. These properties include Li-superionic conductivity\cite{R6}, long-range antiferromagnetic ordering\cite{R7},\cite{R8}, electrocatalysis\cite{R9}, colossal dielectric constant\cite{R10}, thermal conductivity\cite{R11}, and so on. The groundbreaking discovery of the first entropy-stabilized rock salt oxide sparked exploration into various compositions and systems featuring high configurational entropy. Some of these systems exist in entropy-stabilized forms within distinct crystal structures, such as fluorite\cite{R3}, pyrochlore\cite{R12}, however, high entropy forms of spinels\cite{R13}, and perovskites\cite{R14} are also presented. Although the first entropy-stabilized oxide has been a great success in inspiring scientists to develop similar entropy-stabilized oxides with different crystal structures and exploring their unique physical properties. Fracchia et al., in their recent work, showed the single-phase rock-salt structure of samples with low to medium configurational entropy, alongside the finding reported by Rost et al., and raised queries concerning the domination of thermodynamic stability of entropy-stabilized rock-salt samples.\cite{R15} Despite the success of entropy-stabilized oxides, the extent to which configurational entropy governs phase stability, especially in systems with medium $\Delta S_{conf}$, remains unclear. Recent reports suggest stability even below the $1.61\,R$ threshold, questioning prior investigations. A systematic study is therefore required to delineate the competing roles of enthalpy and entropy across varying compositional complexities in rock-salt oxides. In the present work, a comprehensive examination is being undertaken to scrutinize the structural, microstructural, and thermal characteristics of rock-salt oxides spanning low, medium, and high $\Delta S_{conf}$ systems. In particular, the number of elements in the rock-salt structured oxide is increased, and the role of $\Delta S_{conf}$ on the thermodynamic phase stability, thermal conductivity ($\kappa$) and thermoelectric figure of merit is discussed. It has been observed that the $\kappa$ of single-phase samples is smaller than that of the decomposed sample showing the importance of local disorder in maintaining minimum $\kappa$ in these samples.  
\\
\section{Experimental Section}
Rock-salt structured oxides with compositions(Ni$_{0.8}$Cu$_{0.2}$)O,(Ni$_{0.6}$Cu$_{0.2}$Zn$_{0.2}$)O, (Ni$_{0.4}$Cu$_{0.2}$Zn$_{0.2}$Co$_{0.2}$)O,(Ni$_{0.2}$Cu$_{0.2}$Zn$_{0.2}$Co$_{0.2}$Mg$_{0.2}$)O, (NiCuZnCoMg)$_{0.9}$Li$_{0.1}$O, (NiCuZnCoMg)$_{0.9}$K$_{0.1}$O, (NiCuZnCoMg)$_{0.9}$Na$_{0.1}$O were synthesized using standard solid-state reaction, where stoichiometric amount of MgO (Alfa Aesar, 99.9\%), Co$_3$O$_4$ (Alfa Aesar, 99.9\%), CuO (Alfa Aesar, 99.9\%), NiO (Alfa Aesar, 99.9\%), ZnO (Alfa Aesar, 99.9\%), Li$_2$CO$_3$ (Alfa Aesar, 99.9\%), Na$_2$CO$_3$ (Alfa Aesar, 99.9\%), and K$_2$CO$_3$ (Alfa Aesar, 99.9\%) were mixed using planetary ball milling in nominal compositions. The mixture was then compressed into a rectangular bar and sintered at 1223\,K for 18 hours, followed by quenching in air. The sintered bar was ground and ball milled to obtain homogeneous powder, which was further consolidated with spark plasma sintering (SPS) using a Dr. Sinter 515S Syntex set-up at 1173\,K for 5 minutes with a heating and cooling rate of 100 K /min under 100 MPa and using graphite molds. The obtained sintered pellets were annealed in air at 1223\,K for 15 hours to avoid possible carbon contamination and reduction of the materials during SPS and quenched.\\ 
Crystallographic structure and phase identification were performed in finely ground samples using a Cu K$_{\alpha}$ ($\lambda$ = 1.5417~\AA) source on a Bruker D8 diffractometer, followed by profile refinement. The optical band gap was obtained from UV-Visible-NIR spectroscopy measurement using an Agilent Cary 5000 UV-Vis-NIR spectrometer in diffuse reflective spectra (DRS) mode. Surface morphology and chemical compositions were confirmed by SEM (SEM-FEG Zeiss Sigma HD) equipped with EDS. The charge states of each element in all the samples were obtained from the X-ray photoelectron spectroscopy measurement. High resolution transmission electron microscope equipped with energy dispersive x-ray spectroscopy (HRTEM-EDS, JEOL $\&$ JEM-F200 (CF-HR)) were used to observed the nanostructure nature and elemental homogeneity of the samples.\\
The sintered pellets were further cut to proper dimensions using a wire saw for further Seebeck coefficient and electrical resistivity measurements. In a standard four-probe configuration, electrical resistivity and Seebeck coefficient were measured using homemade instruments over a wide temperature range of 300\,K-1150\,K under air. The thermal conductivity ($\kappa$) value was obtained using the following equation: $\kappa$ = $D \cdot \rho_d \cdot C_p$. Thermal diffusivity, $D$, was measured directly in the 300-1073\,K range using a laser flash analyzer (LFA) on a Netzsch LFA-457. Cylindrical pellets of 10\,mm in diameter and 1-2\,mm thick were used for diffusivity measurements. The specific heat capacity ($C_p$) was calculated using the Dulong-Petit law, and the bulk density ($\rho_d$) was calculated using the mass of the sample and its geometric volume. The measured densities of all the samples lie between 85-95\% of the theoretical density.\\
\section{Results and Discussion} 
\begin{figure*}
\centering
  \includegraphics[width=0.95\linewidth]{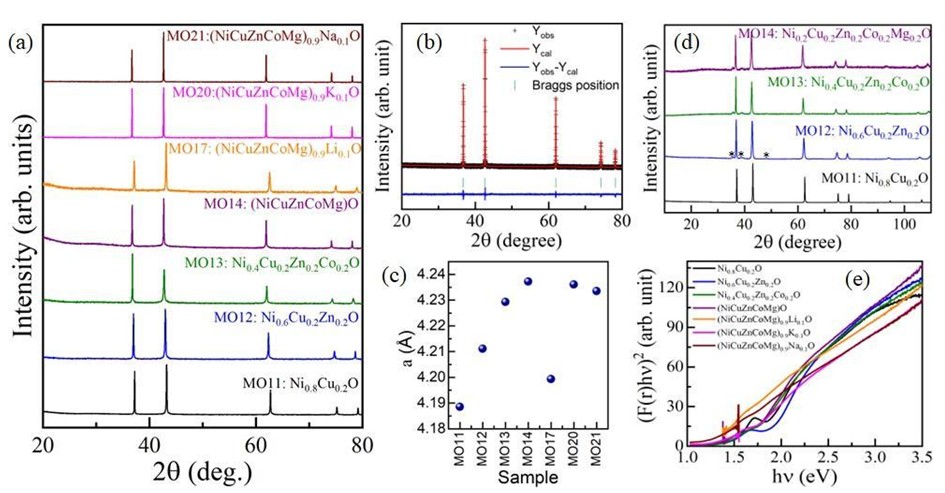}
  \caption{(a) X-ray diffraction pattern for rock-salt structure oxide, (b) Rietveld refinement profile for Li-doped (NiCuZnCoMg)0.9Li0.1O (MO17), confirming phase purity and rock-salt structure, (c) Lattice parameters as a function of composition, showing expansion with increasing cation diversity and alkali doping, (d) X-ray diffraction pattern of the rock-salt samples decomposed at 600$^{\circ}$C, and (e) the Tauc plot obtained from the UV-visible spectroscopy measurement showing direct optical band gaps between 1.25–1.35 eV across all compositions.}
  \label{fig1}
\end{figure*}

\textbf{X-ray Diffraction}: X-ray diffraction pattern for all samples after spark plasma sintering is shown in Fig.~\ref{fig1}(a). All samples have a rock-salt structure, which is in agreement with the report shown by Fracchia et al.\cite{R15} Furthermore, doping with alkali elements (Li$^{+}$, Na$^{+}$, K$^{+}$) also shows a similar crystal structure as shown by Berardan et al.\cite{R6} It is noted that the relative intensity of the Bragg peaks is different with changes in composition. Except for the Li-containing samples, no pattern corresponds to a perfect rock-salt structure. The most striking case is that of MO13:(Ni$_{0.4}$Cu$_{0.2}$Zn$_{0.2}$Co$_{0.2}$)O, as the maximum intensity of the (111) peak is larger than the (200), which constitutes a clear signature of a large structural distortion (it corresponds to a Jahn-Teller distortion around Cu$^{2+}$, which leads to a collective tetragonal distortion\cite{R17}), and this distortion shows a large influence on the transport properties especially these properties are extremely sensitive to the quenching speed and to thermal treatments at temperatures as low as 150$^{\circ}$C.  
Profile matching refinement of all the samples is performed using Fullprof software; the refinement pattern for the Li-containing sample is shown in Fig.~\ref{fig1}(b), and with K$^+$ and Na$^+$ doping is shown in Fig.~S1. and the lattice parameter obtained is shown in Table~\ref{tab:lattice_entropy}. The lattice parameter for the Na-doped sample is larger than that of the K-doped sample Fig.~\ref{fig1}(c). Although potassium (K$^+$) has a larger ionic radius (1.38 \AA) than sodium (Na$^+$, 1.02~\AA), the Na-doped sample exhibits a larger lattice parameter compared to the K-doped sample. This counterintuitive behavior can be attributed to differences in the local structural distortions, bonding environment, and dopant-host interactions.\\
\begin{table*}
    \centering
    \caption{Lattice parameter (a), configurational entropy ($\Delta S$), phase stability, thermal conductivity ($\kappa$) and optical band gap for rock-salt oxides. }
    \label{tab:lattice_entropy}
    \begin{tabular}{lcccccc}
        \hline
        Composition & $a$ (\AA) & $r_{\mathrm{av}}$ (\AA) & $\Delta S_{\mathrm{conf}}$ ($R$) & Entropy stabilized? & $\kappa$ (W\,m$^{-1}$\,K$^{-1}$) & $E_{g}$ (eV) \\
        \hline
        Ni$_{0.8}$Cu$_{0.2}$O & 4.188 & 0.698 & 0.50 & No  & 5.88 & 1.35 \\
        Ni$_{0.6}$Cu$_{0.2}$Zn$_{0.2}$O & 4.211 & 0.708 & 0.95 & Yes & 3.48 & 1.31 \\
        Ni$_{0.4}$Cu$_{0.2}$Zn$_{0.2}$Co$_{0.2}$O & 4.229 & 0.719 & 1.33 & Yes & 3.42 & 1.29 \\
        Ni$_{0.2}$Cu$_{0.2}$Zn$_{0.2}$Co$_{0.2}$Mg$_{0.2}$O & 4.237 & 0.725 & 1.61 & Yes & 2.75 & 1.35 \\
        (NiCuZnCoMg)$_{0.9}$Li$_{0.1}$O & 4.199 & 0.728 & 1.77 & Yes & 2.49 & 1.33 \\
        (NiCuZnCoMg)$_{0.9}$Na$_{0.1}$O & 4.236 & 0.754 & 1.77 & Yes & 2.43 & 1.25 \\
        (NiCuZnCoMg)$_{0.9}$K$_{0.1}$O & 4.233 & 0.790 & 1.77 & Yes & 1.96 & 1.28 \\
        \hline
    \end{tabular}
\end{table*}
\begin{figure*}
\centering
  \includegraphics[width=0.9\linewidth]{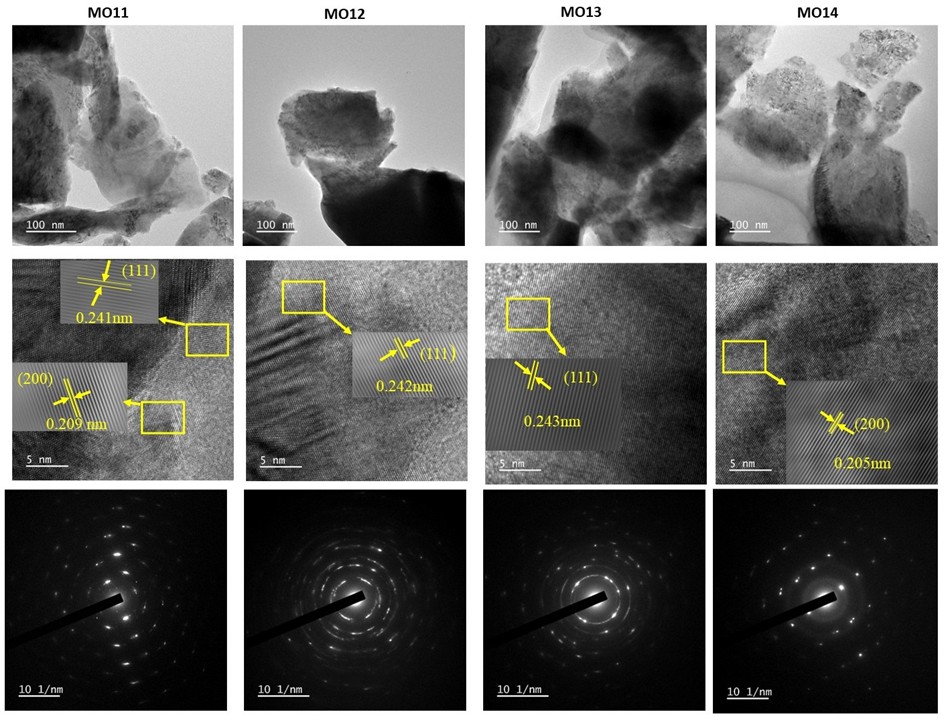}
  \caption{High-resolution transmission electron microscopy (HRTEM) images and selected-area electron diffraction (SAED) patterns for MO11–MO14. Top row: low-magnification images reveal nanocrystalline domains (50–150 nm). Middle row: HRTEM fringes confirm crystalline rock-salt planes with local lattice distortions. Bottom row: SAED patterns display cubic diffraction rings with diffuse scattering, consistent with entropy-driven disorder.}
  \label{fig2}
\end{figure*}
Further, the role of $\Delta S_{conf}$ on thermodynamic phase stability was studied for these compositions, where the increase in the number of cations results in a change from low to medium to high $\Delta S_{conf}$. Fig.~\ref{fig1}(d) shows the XRD pattern of rock salt structured samples decomposed at 600$^{\circ}$C for 6 hours. The sample with three cations, (Ni$_{0.6}$Cu$_{0.2}$Zn$_{0.2}$)O, shows secondary phases when decomposed at 600$^{\circ}$C, which is reversed back to the single phase when further heat treated at 900${^\circ}$C followed by quenching. The same result is observed for the samples with 4 cations:(Ni$_{0.4}$Cu$_{0.2}$Zn$_{0.2}$Co$_{0.2}$)O and 5 cations:(Ni$_{0.2}$Cu$_{0.2}$Zn$_{0.2}$Co$_{0.2}$Mg$_{0.2}$)O, and confirming the thermodynamic stabilization of these compositions by the dominating role of $\Delta S_{conf}$, as mentioned in Table I. Hence, even medium configurational entropy (0.95\,R) for Ni$_{0.6}$Cu$_{0.2}$Zn$_{0.2}$O is significant enough to produce an entropy-stabilized phase in rock-salt structure, which was not answered in studies by Fracchia et al. The stability of the two-cation compound (Ni$_{0.8}$Cu$_{0.2}$)O agrees with the literature reporting the formation of a solid solution up to X$_{Cu}$/(X$_{Cu}$+X$_{Ni}$)~0.25 at 1000$^{\circ}$C.\cite{R16}\\
Further, the optical band gap for all the samples is estimated using UV-Vis-NIR spectroscopy measurement, as shown in Fig.~\ref{fig1}(e). The optical band gap obtained from UV-Vis-NIR measurement is in the range of 1.2-1.3 eV. No change in the band gap (1.3\,eV) is observed by adding K$^+$ in (Ni$_{0.2}$Cu$_{0.2}$Zn$_{0.2}$Co$_{0.2}$Mg$_{0.2}$)O. However, by addition of Li$^+$ or Na$^+$, the optical band gap decreases to 1.25 eV.\\ 
\begin{figure*}
    \centering
    \includegraphics[width=1\linewidth]{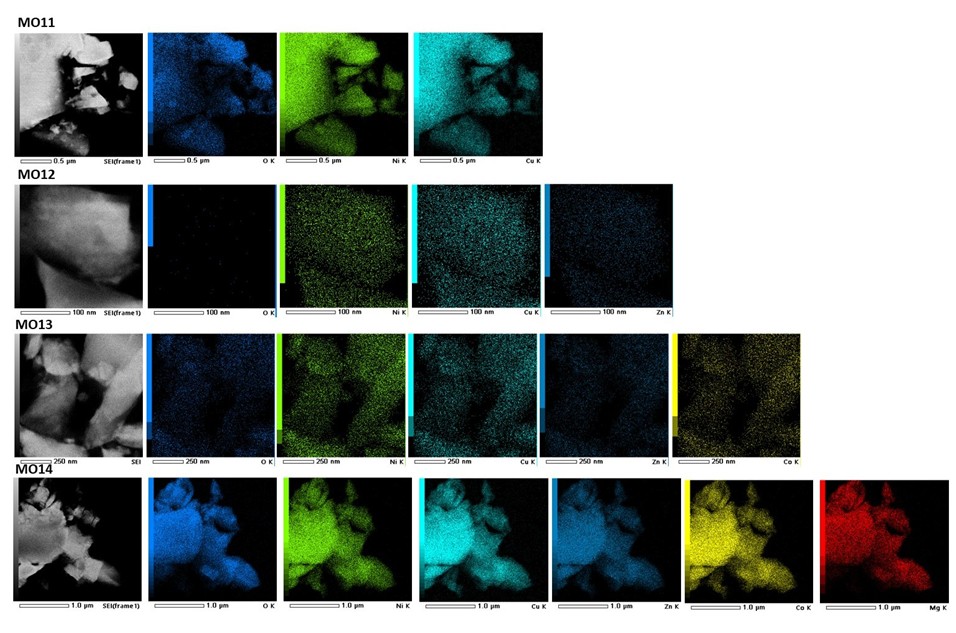}
    \caption{TEM–EDS elemental mapping of rock-salt oxides MO11–MO14. All systems show homogeneous nanoscale distribution of constituent elements (Ni, Cu, Zn, Co, Mg, O), confirming solid-solution formation without clustering or secondary phases. Increasing cationic complexity maintains compositional uniformity, highlighting entropy-driven mixing.}
    \label{fig3}
\end{figure*}
\begin{figure}
    \centering
    \includegraphics[width=1\linewidth]{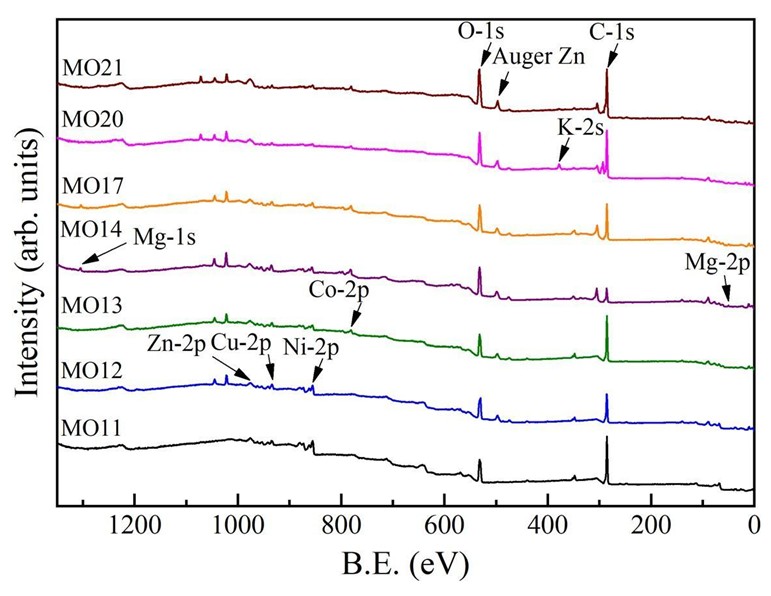} 
    \caption{Survey X-ray photoelectron spectroscopy (XPS) spectra of representative rock-salt oxides (MO11, MO12, MO13, MO14, MO17, MO20, MO21). All expected elements (Ni, Cu, Zn, Co, Mg, O) are detected, with no extraneous peaks, confirming chemical incorporation of designed cations.}
    \label{fig.4}
\end{figure}
\begin{figure*}
    \centering
    \includegraphics[width=0.9\linewidth]{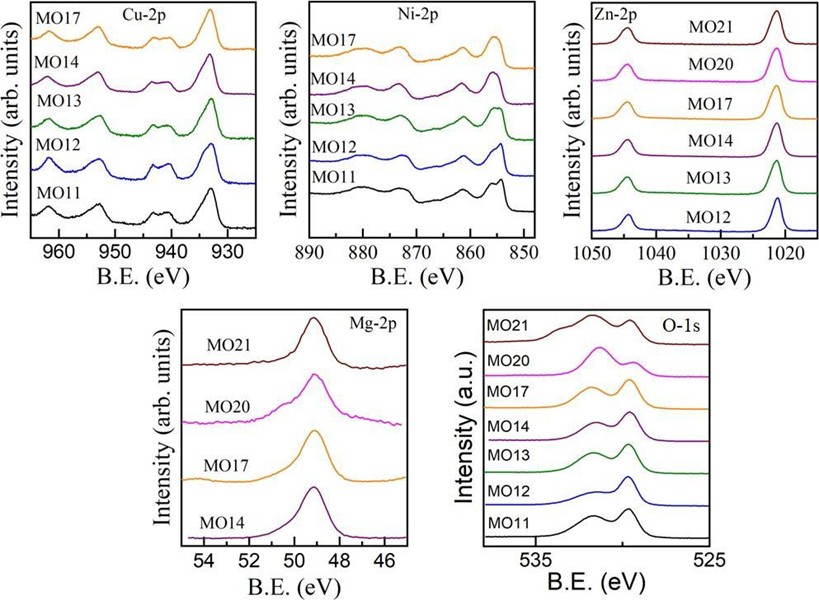} 
    \caption{High-resolution XPS spectra of Cu 2p, Ni 2p, Zn 2p, Mg 2p, and O 1s core levels in rock-salt oxides. Spectra confirm divalent oxidation states of cations: Cu$^{2+}$ (shake-up satellites), Ni2+ (multiplet splitting), Zn$^{2+}$ (sharp 2p doublet), Mg$^{2+}$ (symmetric 2p peak), and lattice O$^{2-}$ with minor vacancy-related components.}
    \label{fig5}
\end{figure*}
\textbf{Field Emission Scanning Electron Microscopy:} Field emission scanning electron microscopy (FE-SEM) equipped with energy-dispersive X-ray spectroscopy shows the homogenous distribution of elements in each sample (Fig.~S3-S4). SEM image for (NiCuZnCoMg)$_{0.9}$Li$_{0.1}$O is shown in Fig.~S6.  Chemical compositions obtained from EDS measurement for each sample are shown in Table S1. Fig.~S3-S8 present the elemental mapping of the synthesized rock-salt oxides obtained via Energy Dispersive X-ray Spectroscopy (EDS), confirming the spatial distribution and homogeneity of constituent elements across the microstructure. The binary sample Ni$_{0.8}$Cu$_{0.2}$O (Fig.~S3) displays uniform distribution of Ni and Cu, supporting its solid solution nature as expected from the literature.\cite{R17a} As the number of cations increases in the ternary and quaternary systems (Fig.~S4-S5), including Zn and Co, the elemental maps continue to show no evidence of segregation, clustering, or phase separation, indicating that these elements are effectively incorporated into the rock-salt matrix. The five-cation high entropy oxide (NiCuZnCoMg)$_{0.9}$Li$_{0.1}$O (Fig.~S6) as well as the alkali-doped analogs with K and Na (Fig.~S7-S8) also exhibit excellent elemental homogeneity at the microscale. The uniform distribution of light elements such as Li and Na, despite the limitations of EDS in detecting low-Z elements, further supports their incorporation, likely at cationic sites or through charge-compensating defect mechanisms\cite{R18}. The absence of secondary phase formation or elemental agglomeration in all compositions confirms that $\Delta S_{conf}$ facilitates the formation of chemically homogeneous, single-phase solid solutions even in highly compositionally complex systems.\\

\begin{figure}
    \centering
    \includegraphics[width=1\linewidth]{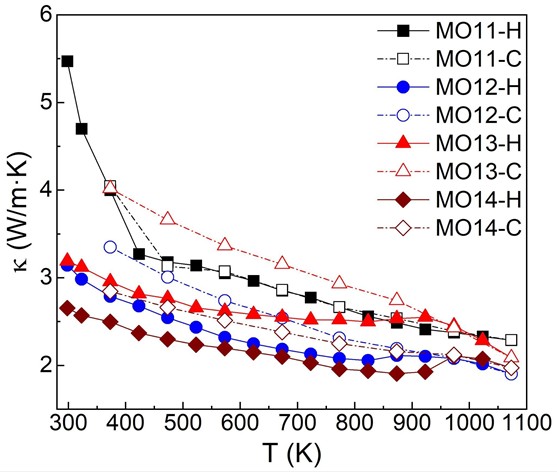} 
    \caption{Thermal conductivity ($\kappa$) as a function of temperature (300–1073 K) for rock-salt oxides under heating (H) and cooling (C) cycles.}
    \label{Fig6*}
\end{figure}

\textbf{High Resolution Transmission Electron Microscopy}: The high-resolution transmission electron microscopy (HRTEM) and selected area electron diffraction (SAED) images of the rock-salt oxides MO11, MO12, MO13, and MO14 are shown in Fig.~\ref{fig2}. The low-magnification TEM micrographs (top row) reveal agglomerated nanocrystalline domains with irregular morphology and particle sizes ranging from 50 to 150 nm, indicative of fine-grained microstructures developed during spark plasma sintering. The HRTEM images (middle row) exhibit well-defined lattice fringes, which confirm the crystalline nature of the samples. The lattice fringes are consistent with interplanar spacings of ~0.241 nm and ~0.209 nm corresponding to the (111) and (200) planes of the rock-salt structure, respectively. Slight variations in fringe contrast and periodicity across the samples suggest local lattice distortions and chemical disorder induced by multicomponent cation mixing. The SAED patterns (bottom row) show concentric diffraction rings with superimposed spot features, indicating the presence of polycrystalline domains with preferred orientation. The diffraction rings correspond to the reflections of (111), (200), (220), and (311) planes of the cubic \textit{Fm-3m} phase, confirming the preservation of the rock-salt structure across all samples. Fig.~\ref{fig3} presents the TEM-EDS elemental mapping of MO11–MO14 samples, confirming uniform distribution of the constituent elements at the nanoscale. No evidence of elemental segregation is observed in any of the samples. These results confirm the successful synthesis of single-phase rock salt oxides.\\ 
\textbf{X-ray Photoelectron Spectroscopy}: To investigate the surface composition and oxidation states of cations in the synthesized rock-salt oxides, X-ray Photoelectron Spectroscopy (XPS) measurements were carried out. The full-range survey spectra (Fig.~\ref{fig.4} confirm the presence of all constituent elements-Ni, Cu, Zn, Co, Mg, and O-in each sample, as per their chemical constituent. The absence of unexpected peaks, along with the well-resolved core levels, indicates phase purity and effective incorporation of all elements into the oxide lattice. The high-resolution core-level spectra (Fig.~\ref{fig5} provide detailed insights into the electronic states of the constituent cations. The Cu-2p spectrum displays prominent satellite peaks beyond the main $2p_{3/2}$ peak ($\sim$934.5~eV), characteristic of the Cu$^{2+}$ oxidation state.\cite{R18a} These satellites originate from shake-up processes associated with the partially filled $3d^{9}$ configuration of Cu$^{2+}$, suggesting strong electron correlation effects.\cite{R18b} The Ni-2p spectrum exhibits broad and asymmetric features centered around 855.5~eV ($2p_{3/2}$), accompanied by multiplet splitting and satellite structures typical of Ni$^{2+}$ in octahedral coordination. The presence of such satellites implies a significant degree of covalency and electron correlation in the Ni–O bonding environment.\cite{R18c} The Zn-2p core level shows sharp peaks at ~1021.7 eV and ~1044.8 eV, corresponding to the$2p_{3/2}$ and $2p_{1/2}$ spin-orbit components, respectively, consistent with the Zn$^{2+}$ oxidation state. The lack of shake-up features supports the closed-shell $d^{10}$ electronic configuration of Zn$^{2+}$.\cite{R18d} The Mg-2p peak, centered near 50.2\,eV, confirms the presence of Mg$^{2+}$ and exhibits a symmetric profile, as expected for a fully occupied shell structure. The O-1s spectrum reveals a dominant peak at ~529.6\,eV attributed to lattice oxygen (O$^{2-}$). In some samples, a shoulder around 531.5\,eV is observed, which can be associated with oxygen vacancies or hydroxyl groups on the surface.\cite{R18e} The relative intensity of this component may reflect variations in local disorder, consistent with the configurational entropy-driven stabilization observed in these materials. Together, the XPS analysis corroborates the divalent oxidation states of the transition and alkaline earth metals in the lattice and highlights the presence of subtle electronic and structural disorder.\\
\textbf{Thermal conductivity Measurement}: The thermal conductivity ($\kappa$) as a function of temperature (T) for the synthesized rock-salt oxides is presented in Fig.~\ref{Fig6*} for both heating and cooling cycles. The data reveal several critical trends that reflect the interplay between $\Delta S_{conf}$, lattice disorder, and phonon scattering mechanisms. At room temperature (300\,K), $\kappa$ of the binary compound Ni$_{0.8}$Cu$_{0.2}$O is the highest ($\sim$5.88~W/m$\cdot$K), consistent with its relatively ordered lattice and lower $\Delta S_{conf}$. As the number of cations increases from 2 to 5, a systematic decrease in $\kappa$ is observed, reaching a minimum of ($\sim$2.75~W/m$\cdot$K) for (Ni$_{0.2}$Cu$_{0.2}$Zn$_{0.2}$Co$_{0.2}$Mg$_{0.2}$)O. This trend is attributed to enhanced phonon scattering arising from increased mass and strain field fluctuations introduced by multi-elemental substitution. The presence of atoms with differing sizes, masses, and bonding characteristics disrupts the phonon transport, effectively reducing the lattice (phonon) thermal conductivity component,$\kappa_{ph}$. Therefore, the decrease in $\kappa$ with the addition of elements may be attributed to the decrease in $\kappa_{ph}$ due to enhanced phonon scattering by lattice distortion and mass fluctuation effects. This effect has already been evidenced in other high-entropy materials.\cite{R19, R20, R21, R22, R23, R24, R25, R26, R27}.\\
The further reduction in thermal conductivity upon substitution with monovalent alkali elements (Li$^+$, Na$^+$, K$^+$) is particularly notable. The larger ionic radii of Na$^+$ and K$^+$ compared to the divalent transition metal cations induce additional local lattice distortions, thereby intensifying phonon scattering. Among these, the K-doped sample shows the lowest $\kappa$ $\sim$2~W/m$\cdot$K, suggesting that lattice softening due to size mismatch is a key driver for $\kappa$ suppression. The temperature dependence of $\kappa$ exhibits a typical decreasing trend with increasing temperature, which is characteristic of phonon-phonon Umklapp scattering dominating the thermal transport.\\
Importantly, the comparison of heating and cooling cycles reveals hysteresis, with $\kappa$ increasing during the cooling cycle, as compared to the value obtained during heating cycle for all the samples who were found to be entropy-stabilized. This increase in $\kappa$ is attributed to the decomposition of entropy-stabilized phases during high-temperature exposure, leading to the formation of secondary phases and reduction in configurational disorder. The resultant partial ordering reduces the scattering of phonons, thus slightly increasing $\kappa$ upon cooling. Thus, the values of $\kappa$ obtained in the present study not only highlight the role of $\Delta S_{conf}$ in stabilizing low-$\kappa$ disordered phases but also demonstrate the tunability of phonon transport through a precise change in $\Delta S_{conf}$.\\

\begin{figure}
    \centering
    \includegraphics[width=0.99\linewidth]{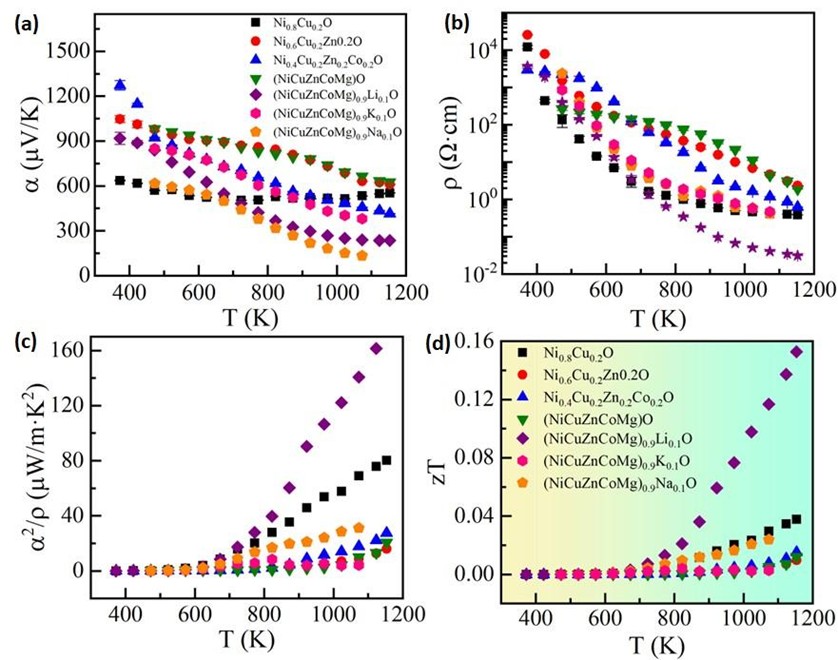}
    \caption{(a) Seebeck coefficient ($\alpha$), (b) electrical resistivity ($\rho$), 
    (c) power factor $\left(\frac{\alpha^{2}}{\rho}\right)$, and (d) figure of merit ($zT$) 
    as a function of temperature for rock-salt samples.}
    \label{fig7}
\end{figure}

\textbf{Seebeck coefficient, Electrical Conductivity and Figure of merit}: Seebeck coefficient ($\alpha$) as a function of temperature (heating cycle) for all samples is shown in Fig.~\ref{fig7}(a).  All samples show a positive value of $\alpha$, depicting the majority of carriers as holes. The samples show a high $\alpha$ close to 400\,K, which decreases with increasing temperature, showing non-degenerate semiconductor behavior. 
Temperature-dependent electrical resistivity ($\rho$ vs T) for all samples is shown in Fig.~\ref{fig7}(b). $\rho(T)$ shows semiconducting behavior with $\rho$ values decreasing with increasing temperature. $\rho$ increases as the number of cations increases in the rock-salt structure compared to (Ni$_{0.8}$Cu$_{0.2}$)O. Such a change in $\rho$ may be attributed to (i) an increased disorder, which may scatter the charge carriers and may reduce carrier mobility, and (ii) a change in the band structure. However, further experimental or theoretical studies are required to confirm this. Now, looking at the effect of alkaline doping, a decrease in $\rho$ and $\alpha$ is observed for samples containing Li, Na, and K compared to the non-doped high entropy materials. At temperatures greater than 700\,K, $\rho$ for (NiCuZnCoMg)$_{0.9}$Li$_{0.1}$O is significantly lower as compared to all other samples. This may be understood as (i) Li$^+$ has similar ionic radii as compared with Ni$^{2+}$, Cu$^{2+}$, Zn$^{2+}$, Co$^{2+}$, and Mg$^{2+}$ and hence induces less lattice distortion which therefore results in minimal charge carrier mobility deterioration, and ii) some ionic Li+ conduction can occur\cite{R6}. The power factors ($\alpha_{2}/\rho$) as a function of temperature for all samples are shown in Fig.~\ref{fig7}(c). It is seen that $\alpha_{2}/\rho$ increases with increasing temperature and is attributed to the decrease in $\rho$ of the samples. A maximum power factor of $160~\mu\mathrm{W}/\mathrm{m\cdot K^2}$ is obtained for Li-doped (NiCuZnCoMg)$_{0.9}$Li$_{0.1}$O.

The figure of merit, $zT$, as a function of temperature is calculated for all samples (using the temperature-dependent values of $\alpha$, $\rho$, and $\kappa$) and is shown in Fig.~\ref{fig7}(d). $zT$ increases monotonically with rising temperature for all samples, and a maximum $zT$ of 0.15 is obtained for (NiCuZnCoMg)$_{0.9}$Li$_{0.1}$O at 1173\,K. This is one of the highest values of $zT$ in rock-salt oxides at this temperature.  Among the alkali elements (Li$^+$, Na$^+$, K$^+$) doping, Li$^+$ is found to be the most effective for enhancing the thermoelectric properties of rock-salt entropy-stabilized oxides. However, ionic conduction could deteriorate the properties, preventing any long-term use. In addition, the entropy-stabilized nature of the sample restricts its practical application at temperatures below 900$^{\circ}$C, as it decomposes at lower temperatures (Fig.~S2).\\
\section {Conclusion}
This study provides a comprehensive investigation of the thermodynamic phase stability and transport behavior of high-entropy rock-salt oxides as a function of configurational entropy ($\Delta S_{conf}$). It is found that by systematically varying the cationic complexity from 2 to 5 elements in rock-salt structure, a single-phase rock salt structure is observed in every sample; however, sample with 3 elements in equimolar ratio shows an entropy-stabilized nature with $\Delta S_{conf}$ as low as 0.95R. XRD measurement of samples with thermal cycling confirms reversible phase transitions in medium to high entropy systems, indicating an entropy-stabilization even below the conventional threshold of $\Delta S_{conf}$~1.61\,R. Field emission scanning electron microscope and transmission electron microscopy equipped with EDS mapping further confirms the chemical homogeneity of the sample at micro and nano level. Thermal conductivity measurements reveal a sharp suppression in $\kappa$ with increasing elemental diversity, reaching $\sim 2~\mathrm{W}/\mathrm{m\cdot K}$ for alkali-doped compositions at 300\,K due to enhanced phonon scattering from mass and strain field fluctuations. The hysteresis observed in thermal transport between heating and cooling cycles shows the role of $\Delta S_{conf}$ in achieving minimum $\kappa$. Thermoelectric measurement performed on these samples shows a promising thermoelectric figure of merit ($zT \sim 0.15$ at 1173\,K), driven by low $\kappa$ and favorable electronic properties in Li$^+$ doped sample.\\
%
%\section*{Author Contributions}
%\textbf{Ashutosh Kumar}: Conceptualization, Investigation, Formal Analysis, Writing – original draft, \textbf{David Berardan}: Conceptualization, Investigation, Formal Analysis, Writing – review & editing. \textbf{Francois Brisset}: Investigation, Writing – review & editing, \textbf{Diana Dragoe}: Investigation, Formal Analysis, Writing – review & editing, \textbf{Nita Dragoe}:  Conceptualization, Visualization, Supervision, Formal Analysis, Funding acquisition, Project administration, Writing – review & editing. 
%
%
\section*{Conflicts of interest}
There are no conflicts to declare.
\section*{Acknowledgements}
This work was supported by the French Agence Nationale de la Recherche (ANR), through the project NEO (ANR 19-CE30-0030-01). J. Kumar and A. Kumar also thank the CIF, IIT Bhilai for HRTEM measurement, \\
%

%%%REFERENCES%%%

%
%
%
\section{Supporting Information}

Rietveld refinement pattern for (a) (NiCuZnCoMg)$_{0.9}$Na$_{0.1}$O and (b) (NiCuZnCoMg)$_{0.9}$K$_{0.1}$O. XRD pattern of the rock-salt oxide samples sintered at 600$^{\circ}$C for 6 hours followed by slow cooling, Energy dispersive X-ray spectroscopy (EDS) mapping for Ni$_{0.8}$Cu$_{0.2}$O, Ni$_{0.6}$Zn$_{0.2}$Cu$_{0.2}$O, Ni$_{0.4}$Co$_{0.2}$Zn$_{0.2}$Cu$_{0.2}$O, ((MgCoNiCuZn)$_{0.9}$Li$_{0.1}$)O, ((MgCoNiCuZn)$_{0.9}$K$_{0.1}$)O, ((MgCoNiCuZn)$_{0.9}$Na$_{0.1}$)O.\\
%
%If notes are included in your references you can change the title from 'References' to 'Notes and references' using the following command:
%\renewcommand\refname{Notes and references}
\newpage
\begin{figure}
    \centering
    \includegraphics[width=\linewidth]{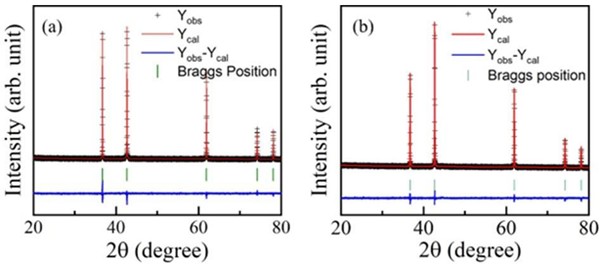} 
    \caption{S1. Rietveld refinement pattern for (a) (NiCuZnCoMg)$_{0.9}$Na$_{0.1}$O and (b) (NiCuZnCoMg)$_{0.9}$K$_{0.1}$O.}
    \label{figS1}
\end{figure}

\begin{figure}
    \centering
    \includegraphics[width=\linewidth]{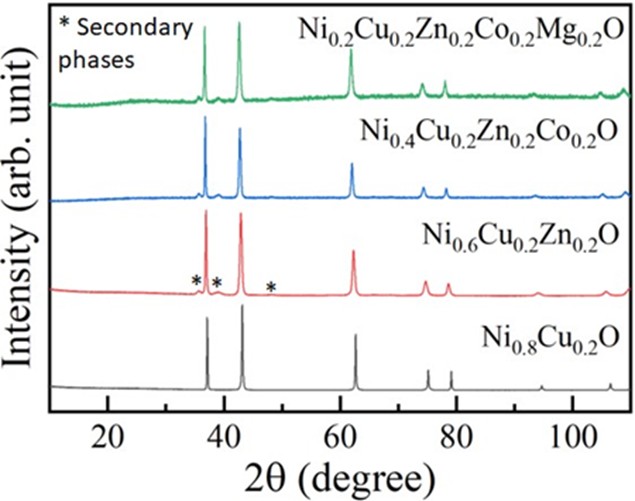}
    \caption{S2. XRD pattern of the rock-salt oxide samples sintered at 600$^{\circ}$C for 6 hours followed by slow cooling.}
    \label{figS2}
\end{figure}

\begin{figure*}
    \centering
    \includegraphics[width=0.95\linewidth]{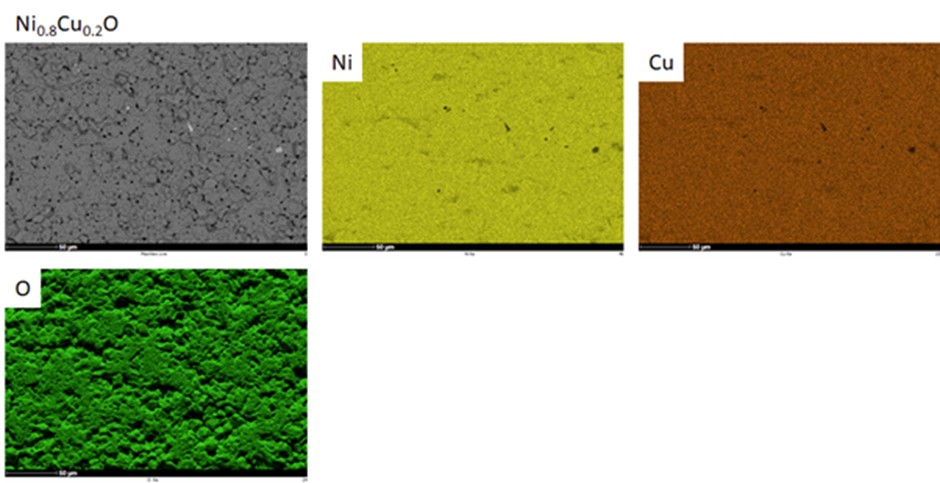}
    \caption{S3. Energy-dispersive X-ray spectroscopy (EDS) mapping for Ni$_{0.8}$Cu$_{0.2}$O.}
    \label{figS3}
\end{figure*}

\begin{figure*}
    \centering
    \includegraphics[width=0.95\linewidth]{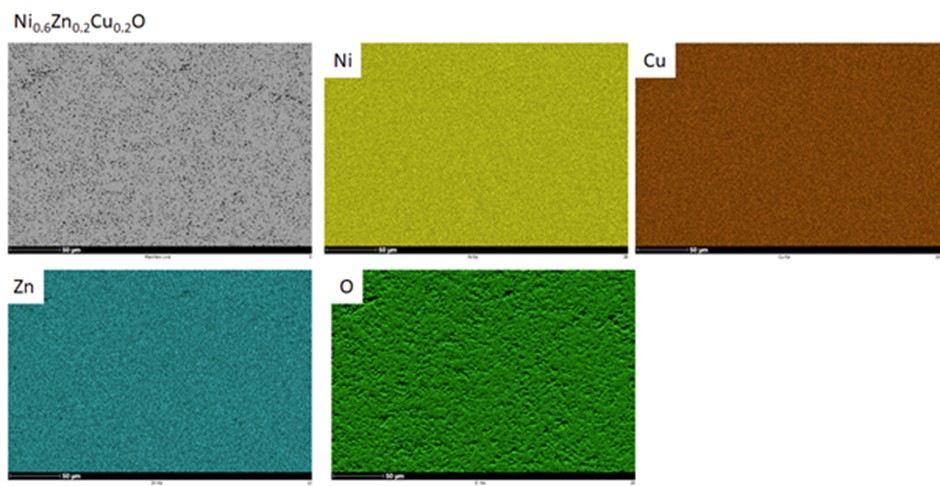} 
    \caption{S4. Energy-dispersive X-ray spectroscopy (EDS) mapping for Ni$_{0.6}$Zn$_{0.2}$Cu$_{0.2}$O.}
    \label{figS4}
\end{figure*}

\begin{figure*}
    \centering
    \includegraphics[width=0.95\linewidth]{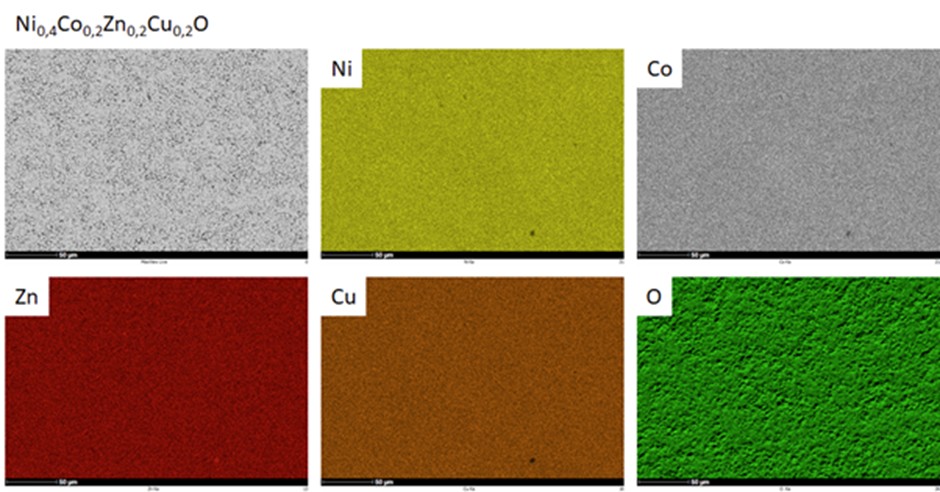} 
    \caption{S5. Energy-dispersive X-ray spectroscopy (EDS) mapping for Ni$_{0.4}$Co$_{0.2}$Zn$_{0.2}$Cu$_{0.2}$O.}
    \label{figS5}
\end{figure*}

\begin{figure*}
    \centering
    \includegraphics[width=0.95\linewidth]{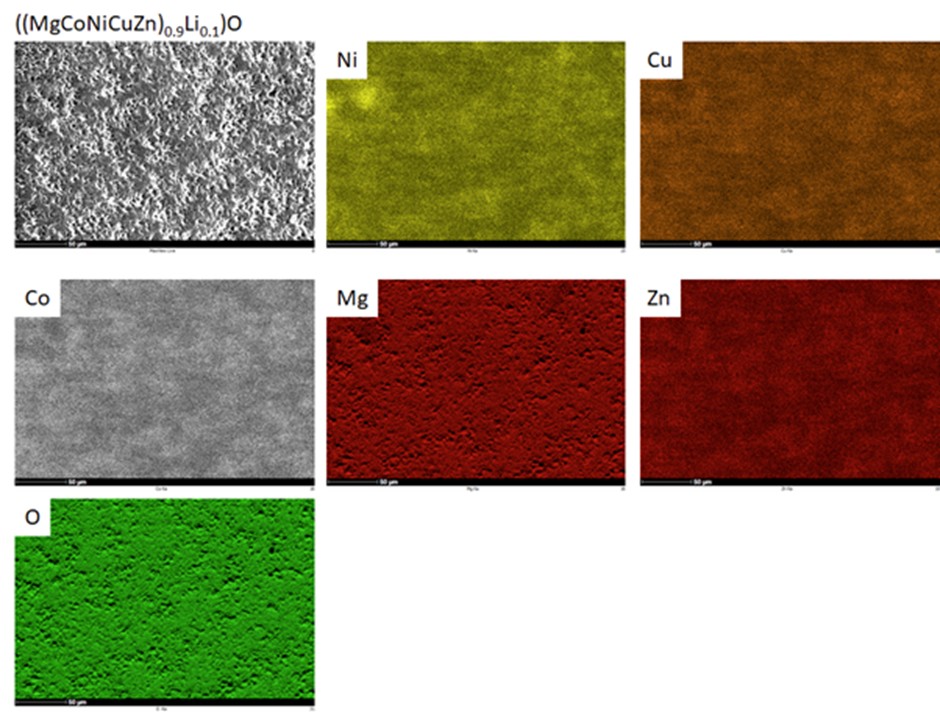} 
    \caption{S6. EDS mapping for ((MgCoNiCuZn)$_{0.9}$Li$_{0.1}$)O.}
    \label{figS6}
\end{figure*}

\begin{figure*}
    \centering
    \includegraphics[width=0.95\linewidth]{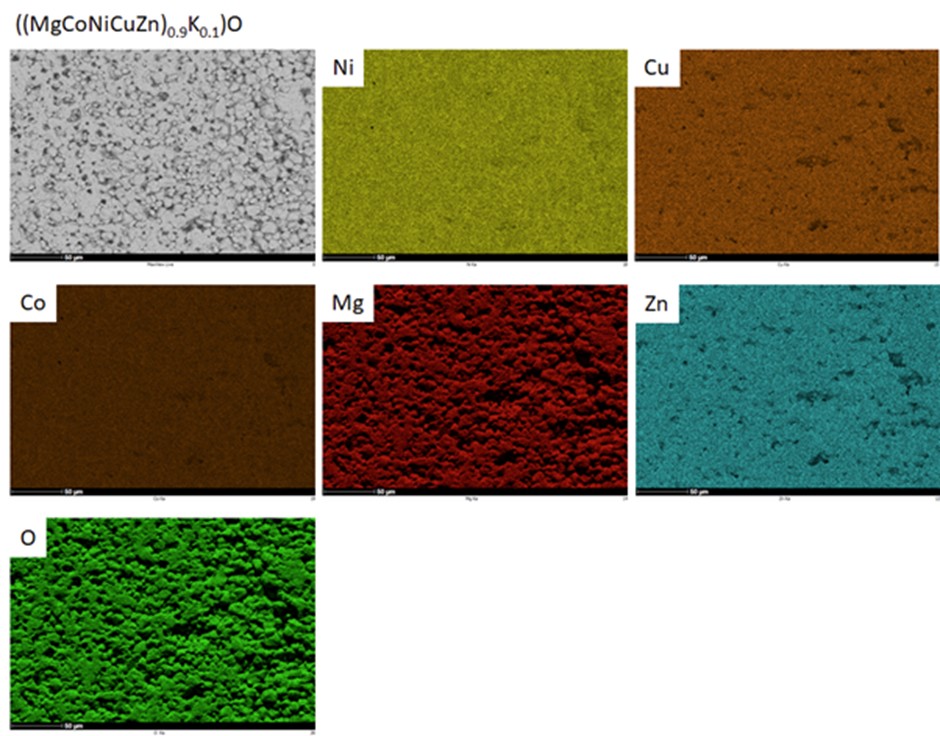} 
    \caption{S7. EDS mapping for ((MgCoNiCuZn)$_{0.9}$K$_{0.1}$)O.}
    \label{figS7}
\end{figure*}

\begin{figure*}
    \centering
    \includegraphics[width=0.95\linewidth]{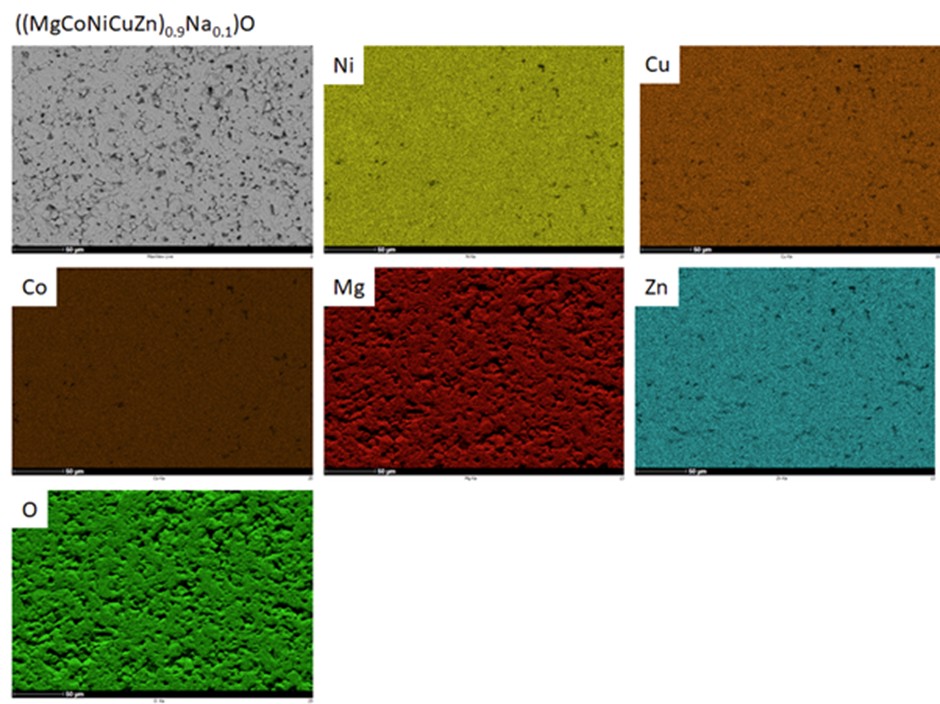} 
    \caption{S8. EDS mapping for ((MgCoNiCuZn)$_{0.9}$Na$_{0.1}$)O.}
    \label{figS8}
\end{figure*}

\begin{table*}
\centering
\caption{Table SI. Chemical composition (at.\%) measured by EDS. The measurement precision is only semi-quantitative at these low values and for oxygen (O). Th = theoretical, Ex = experimental, ND = not detectable.}
\label{tab:SI_eds}
\begin{tabular}{c c c c c c c c c c c}
\hline
\textbf{Sample} & \textbf{Ni} & \textbf{Cu} & \textbf{Zn} & \textbf{Co} & \textbf{Mg} & \textbf{A (Li/Na/K)} & \textbf{O} & \textbf{Source} \\
\hline
Ni$_{0.8}$Cu$_{0.2}$O & 50 &40 & 10 & 0.00 & 0.00 & 0.00 & 0.00 & Th\\
Ni$_{0.8}$Cu$_{0.2}$O & 49.5 &40.2 &10.3 &0.00 &0.00 & 0.00 & 0.00 & Ex\\
Ni$_{0.6}$Cu$_{0.2}$Zn$_{0.2}$O &5 &3 &10 &10 &0.00 &0.00 &0.00 &0.00 &0.00 & Th \\
Ni$_{0.6}$Cu$_{0.2}$Zn$_{0.2}$O &46.8 &21.9 &11.1 &10.1 &0.00 &0.00 &0.00 &0.00  &0.00 & Ex \\
Ni$_{0.4}$Cu$_{0.2}$Zn$_{0.2}$Co$_{0.2}$O &50 &20 &10 &10 &10 &0.00 &0.00 &0.00 &0.00 & Th \\
Ni$_{0.4}$Cu$_{0.2}$Zn$_{0.2}$Co$_{0.2}$O &49.5 &20.3 &10.2 &9.7 &10.3 &0.00 &0.00 &0.00 &0.00 & Ex \\
Ni$_{0.2}$Cu$_{0.2}$Zn$_{0.2}$Co$_{0.2}$Mg$_{0.2}$O &50 &10 &10 &10 &10 & 10 &0.00 &0.00 &0.00 & Th \\
Ni$_{0.2}$Cu$_{0.2}$Zn$_{0.2}$Co$_{0.2}$Mg$_{0.2}$O &57.0 &7.8 &7.5 &7.1 &8.1 &12.6 &0.00 &0.00 &0.00 & Ex \\
NiCuZnCoMg$_{0.9}$Li$_{0.1}$O &50 &9 &9 &9 &9 &9 &5 &0.00 &0.00 & Th \\
NiCuZnCoMg$_{0.9}$Li$_{0.1}$O &50.9 &9.9 &9.0 &9.8 &7.6 &12.8 &ND &0.00 &0.00 & Ex \\
NiCuZnCoMg$_{0.9}$Na$_{0.1}$O &50 &9 &9 &9 &9 &9  &0.00 &0.00 &5  & Th \\
NiCuZnCoMg$_{0.9}$Na$_{0.1}$O &46.3 &9.7 &9.2 &9.1 &9.6 &10.6 &0.00 &0.00 &5.5 & Ex \\
NiCuZnCoMg$_{0.9}$K$_{0.1}$O  &50 &9 &9 &9 &9 &9 &0.00 &5 &0.00 & Th \\
NiCuZnCoMg$_{0.9}$K$_{0.1}$O  &41.1 &10.6 &10.0 &9.6 &10.5 &11.2 &0.00 &0 &0.00 & Ex \\
\hline
\end{tabular}
\end{table*}

%\bibliography{main} %You need to replace "rsc" on this line with the name of your .bib file
%\bibliographystyle{rsc} %the RSC's .bst file
\end{document}